\documentclass[aps,prl,twocolumn,groupedaddress,superscriptaddress,showpacs,floatfix]{revtex4-1}
\usepackage{graphicx}

\usepackage{xcolor}
\usepackage{amsfonts}
\usepackage{amsmath}
\usepackage{amssymb}
\usepackage{rotating}
\usepackage{bm}
\usepackage{lipsum}
\usepackage{bbm}

\usepackage[unicode=true, bookmarks=false,breaklinks=false,pdfborder={0 0 1},backref=false,colorlinks=false]{hyperref}
\hypersetup{hypertex}   % gera link para referencias ao longo do texto
\usepackage{breakurl}
\hypersetup{backref,colorlinks=true,citecolor=blue,linkcolor=blue,filecolor=blue,urlcolor=blue}
\usepackage[english]{babel}

\begin{document}

\title{Spin-polarization control driven by a Rashba-type effect breaking the mirror symmetry in two-dimensional dual topological insulators}

\author{Carlos Mera Acosta}
\email[]{acosta@if.usp.br}
\affiliation{Institute of Physics, University of Sao Paulo, CP 66318, 05315-970, Sao Paulo, SP, Brazil}
\affiliation{Brazilian Nanotechnology National Laboratory, CP 6192, 13083-970, Campinas, SP, Brazil}

\author{Adalberto Fazzio} 
\affiliation{Institute of Physics, University of Sao Paulo, CP 66318, 05315-970, Sao Paulo, SP, Brazil}
\affiliation{Brazilian Nanotechnology National Laboratory, CP 6192, 13083-970, Campinas, SP, Brazil}

%\date{\today}

\begin{abstract}
Three-dimensional topological insulators protected by both the time reversal (TR) and mirror symmetries were recently predicted and observed. Two-dimensional materials featuring this property and their potential for device applications have been less explored. %In these systems, the edge states spin-polarization is oriented perpendicular to the mirror plane that protects them.
%Using density functional theory calculations and a tight-binding effective model, 
%In this work, we find that in dual topological insulator, an external electric field breaking the mirror symmetry induces a spin-polarization parallel to the mirror plane in the time reversal protected edge states.
We find that in these systems, the spin-polarization of edge states can be controlled with an external electric field breaking the mirror symmetry. This symmetry requires that the spin-polarization is perpendicular to the mirror plane, therefore, the electric field induces spin-polarization components parallel to the mirror plane.
Since this field preserves the TR topological protection, we propose a transistor model using the spin-direction of protected edge states as a switch.
In order to illustrate the generality of the proposed phenomena, we consider compounds protected by mirror planes parallel and perpendicular to the structure, e.g., Na$_3$Bi and half-functionalized (HF) hexagonal compounds, respectively. For this purpose, we first construct a tight-binding effective model for the Na$_3$Bi compound and predict that HF-honeycomb lattice materials are also dual topological insulators.
%We then show in a general way that the spin-polarization can be controlled by applying out-plane and in-plane external electric fields in Na$3$Bi and half-functionalized hexagonal compounds, respectively.
\end{abstract}

\maketitle

The quantum geometrical description of the insulator state gave rise to a breakthrough in the understanding of the topological phases in solids~\cite{TKNN,Resta2011,Resta1993,Hasan2012}. % of matter.  
Topological invariants,~e.g.,~the $Z_{2}$-invariant and the Chern number $\mathcal{C}_{n}$, classify insulators according to the preserved symmetries and the "symmetry-charge" pumped to the boundary~\cite{Fu2006}. Systems featuring a non-zero topological invariant,~i.e.,~topological insulators (TIs), support dissipationless metallic boundary (edge/surface) states protected by a specific crystal symmetry on a bulk insulator~\cite{Qi2011}.
For instance, quantum spin Hall insulators (QSHIs) and topological crystalline insulators (TCIs)  are two-dimensional (2D) materials characterized by $Z_{2}=1$~\cite{PhysRevLett.95.226801,Kane2005,Fu2007a} and a non-zero mirror Chern number~\cite{PhysRevLett.106.106802,ncomms1969,YoichiAndo_LiangFu2015}, $\mathcal{C}_{M}$, respectively. 
In QSHIs, the edge states are protected by the time-reversal (TR) symmetry, while in TCIs by either point or mirror symmetries.
Topological transitions are typically related to band inversions~\cite{RevModPhys_Basil}: the transition from normal insulators to either QSHIs or TCIs with $\mathcal{C}_{M}=\pm 1$ requires an odd number of band inversions, while TCIs with $\mathcal{C}_{M}=\pm 2$ exhibit even band inversions. Naturally, $\mathcal{C}_{M}\neq 0$ or $\pm 2$ intrinsically avoid the QSH state ($Z_{2}=1$). 

Materials with a dual topological character (DTC),~i.e.,~systems that are simultaneously QSHIs (TIs in three-dimensions) and TCIs~\cite{PhysRevLett.112.016802}, requires both $Z_{2}=1$ and $\mathcal{C}_{M}=\pm 1$, which imposes a condition: it must occur only an odd number of band inversion at $k$-points preserving both the mirror and the TR symmetries. 
In view of the TR invariant momentum points always come in pairs, this condition is only satisfied by the $\Gamma$ point.    
%Thus a natural way to search for materials featuring DTC is to look for TCIs among the already predicted QSHIs in which the band inversion takes place at the $\Gamma$ point.
Since 2D-TCIs are typically systems with $\mathcal{C}_{M}=\pm 2$,~e.g.,~SnTe multilayers~\cite{ncomms1969,Ando2012}, monolayers of SnSe, PbTe, PbSe~\cite{Tome2014,Liu_Liang2015,PhysRevB.91.201401}, TlSe~\cite{Mokrousov2015}, and SnTe/NaCl quantum wells~\cite{Mokrousov2016}, DTC in 2D compounds have been predicted only for Na$_3$Bi layers~\cite{Na3Bi_2017}. % and graphene~\cite{Liang2013,PhysRevLett.95.226801}. 
Three dimensional materials exhibiting DTC not only have been predicted in Bi$_{2}$Te$_{3}$~\cite{PhysRevLett.112.016802}, Bi$_{4}$Se$_{3}$~\cite{PhysRevLett.114.256401}, and Bi$_{1-x}$ Sb$_{x}$~\cite{PhysRevB.78.045426}, but also experimentally observed in the stoichiometric superlattice [Bi$_2$]$_{1}$[Bi$_2$Te$_3$]$_{2}$~\cite{Bi1Te1}. 
%Since this effect allows the spin-polarization control, 

The search for novel systems featuring DTC and the study of their potential for device applications is desired for the development of spintronics.
For instance, in systems featuring DTC external, electric and magnetic fields %perpendicular to the mirror plane 
allow the control of topological states.
This intrinsic property arises from the possible separate manipulation of QSHI~\cite{Qian1344,PhysRevLett.109.076601,PhysRevB.96.155302} and TCIs~\cite{Liang2013,Okada1496,Dziawa2012} states due to the symmetry breaking induced by the external fields, i.e., the magnetic (electric) field breaks the TR (mirror) symmetry, but the edge states could still be protected by the mirror (TR) symmetry.
%Naturally, the field effects also include the orbital- and spin-texture manipulation, e.g., spin-polarization control, induced magnetization and Rashba spin-splitting, and ban-gap size engineering.

\begin{figure}
  \centering
  \includegraphics[width = 0.95 \linewidth]{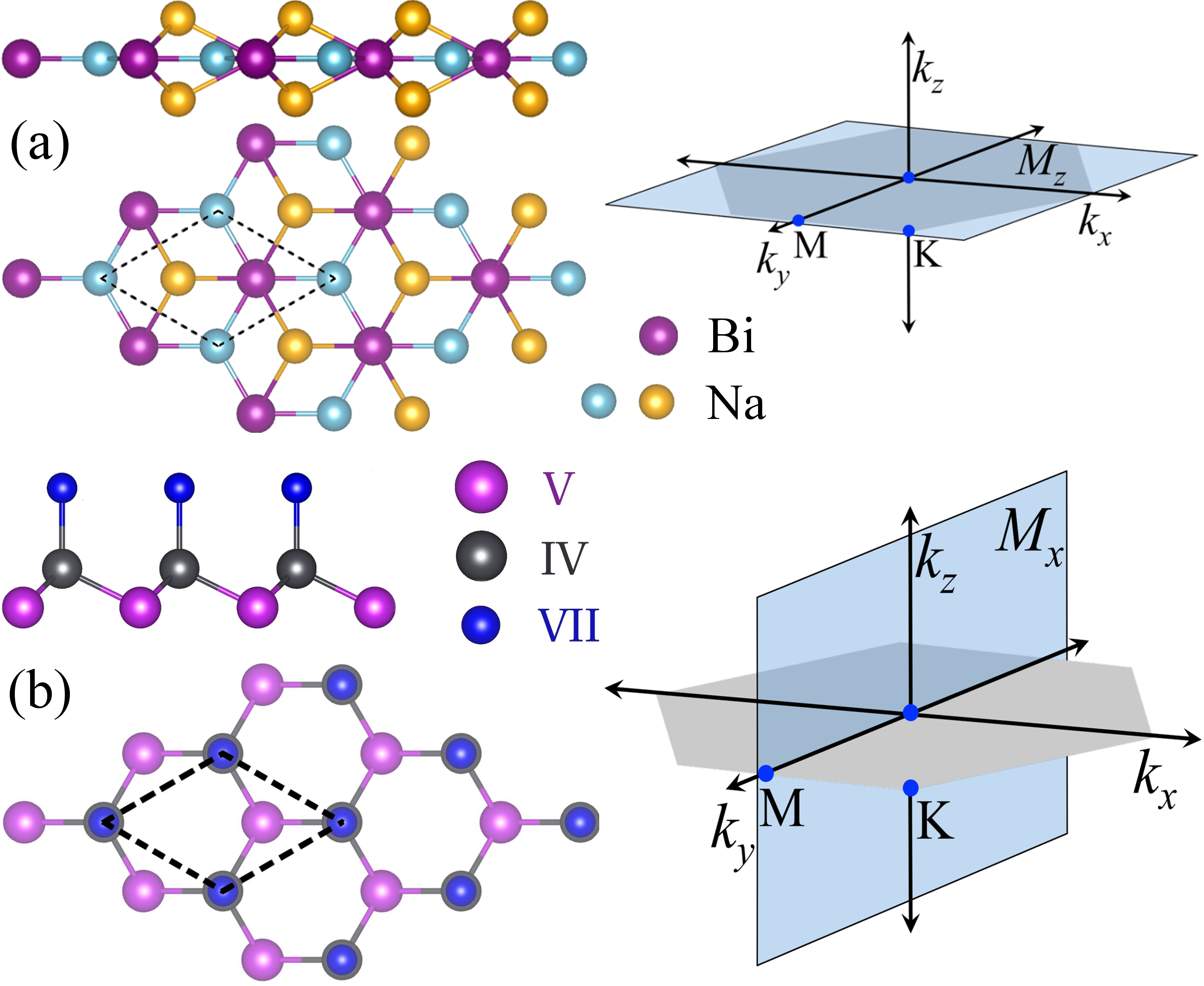}  
  \caption{Side and top view of the (a) Na$_3$Bi and (b) HF honeycomb lattice compounds. Unit cells are represented by the dashed lines and the respective Brillouin Zone (BZ) in gray. The mirror planes given rise to the TCI state are represented by the blue planes. For HF compounds, there are three equivalent reflection planes containing the line connecting two TR invariant equivalent M-points. For illustrative purpose, we only show the plane giving the reflection $x\rightarrow -x$.}
 \label{f:Fig1}
\end{figure}

In this letter, we propose the control of the edge states spin-polarization orientation through an external electric field breaking the mirror symmetry in systems exhibiting DTC, i.e., this field induced a Rashba-type effect resulting in spin-polarization components parallel to the mirror plane. 
Based on this effect, we suggest a transistor model using the spin-direction as a switch.
In order to illustrate the generality of the proposed phenomena, we consider compounds protected by mirror planes parallel and perpendicular to the structure, e.g., Na$_3$Bi and half-functionalized (HF) hexagonal compounds, respectively. For this purpose, we first \textit{i}) construct a tight-binding effective model for the Na$_3$Bi compound~\cite{Na3Bi_2017}, and \textit{ii}) predict that the TR-symmetry protected edges states in HF-honeycomb lattice materials Ref.~\onlinecite{MeraAcosta2016} are also protected by mirror planes perpendicular to the layer, which is different from the conventional TCIs. 
Using the proposed models, we then study the in- and out-plane electric field breaking the mirror symmetry in Na$_3$Bi and HF-hexagonal compounds, verifying the change of the spin-polarization in TR-symmetry protected edge states.

\textit{Effective Hamiltonian describing the DTC of Na$_{3}$Bi:} 
The Na$_{3}$Bi compound satisfies the symmetry operations: \textit{i}) three-fold rotation symmetry $\mathcal{R}_3$ along the $z$-axis, \textit{ii}) TR symmetry $\mathcal{T}$, \textit{iii}) mirror symmetry $\mathcal{M}_{x}$ ($x\to -x$), and 
\textit{iv}) mirror symmetry $\mathcal{M}_{z}$ ($z\to -z$). The TC phase is protected by the reflection symmetry respect to the plane containing the 2D layer, i.e., $M_{z}$, which is schematically represented in Fig \ref{f:Fig1}.
%the two triangular sublattices forming the honeycomb structure.
In the inverted order, the valence band maximum (VBM) has a $p$-orbital character, mainly dominated by $p_{z}$- and $s$-orbitals of the Bi atoms, whereas the conduction band minimum (CBm) mainly consists of $p_{x,y}$-Bi orbitals.
At the $\Gamma$ point, the valence (conduction) band is described by the effective states $\{|\textrm{Bi}_{J},j_{z}\rangle\}$ with the total angular momentum $J=3/2$ ($J=1/2$), i.e, $|\textrm{Bi}_{1/2},\pm 1/2\rangle$ and $|\textrm{Bi}_{3/2},\pm 1/2\rangle$, respectively.  
Using the symmetry operations for these states, we find an effective tight-binding Hamiltonian model describing the Na$_{3}$Bi band structure near the Fermi energy (See supplemental material). In the limit $\vec{k}\rightarrow\Gamma$, the effective Hamiltonian reads
\begin{equation}
\mathcal{H}_{Na_{3}Bi}=\left(\begin{array}{cccc}
-\varepsilon +\xi k^{2} & 0 & 0 & \ell_{-}(\vec{k}) \\
0 & -\varepsilon +\xi k^{2}& \ell_{+}(\vec{k}) & 0\\
0 &  \ell_{-}(\vec{k}) &  \tilde{\varepsilon} -\tilde{\xi}  k^{2} & 0 \\
 \ell_{+}(\vec{k}) & 0 &0 & \tilde{\varepsilon} - \tilde{\xi} k^{2}
\end{array}\right),
\label{HNa3Bi}
\end{equation}
where $k_{\pm}=k_{x}\pm ik_{y}$ and $k_{2}=k_{x}^{2}+k_{y}^{2}$. Here, $\ell{\pm}=\gamma k_{\pm} + \tilde{\gamma}k^{2}_{\mp}$ is the interaction term between the $J=1/2$ and $J=3/2$ states. 
The on-site energy (mass term) and the kinetic term for states with $J=1/2$ ($J=3/2$) are represented by $\varepsilon$ and $\xi$ ($\tilde{\varepsilon}$ and $\tilde{\xi}$), respectively. This model reproduces our density functional theory (DFT) calculation, performed using the SIESTA code\cite{soler2002siesta} with the on-site approximation for the SOC\cite{PhysRevB.89.155438,fernandez2006site} and the Perdew-Burke-Ernzenhof generalized gradient approximation\cite{PhysRevLett.77.3865} for the exchange-correlation functional.

%The band inversion is then introduced only by considering different signs in these terms, i.e., the inverted band-gap at the $\Gamma$ point is defined by the difference between the mass terms,~i.e.,~$E_{g}(\vec{k}\rightarrow\Gamma)=\tilde{\varepsilon}+\varepsilon$.

The mirror symmetry protection becomes evident when the Hamiltonian is written in the basis formed by the mirror operator eigenvectors. 
For any generic $\vec{k}$-point in the BZ, the wave-function $|\psi_{n}^{\sigma}(k_{x},k_{y})\rangle$ (with $\sigma=\uparrow,\downarrow$) can be indexed with the eigenvalues of the mirror operator $\mathcal{M}_{z}$, $m_{z,J}$.
This is a consequence of commutation relation $[\mathcal{H}(k_{x},k_{y}),\mathcal{M}_{z}]$, which can be easily verified using the  matrix representation $\mathcal{M}_{z}=-i\tau_{z}\otimes\sigma_{z}$.,~i.e.,~the mirror operator $\mathcal{M}_{z}$ transforms the orbitals as $\mathcal{M}_{z}|\textrm{Bi}_{1/2},\pm 1/2\rangle=\mp i|\textrm{Bi}_{1/2},\pm 1/2\rangle$ and $\mathcal{M}_{z}|\textrm{Bi}_{3/2},\pm 1/2\rangle=\pm i|\textrm{Bi}_{3/2},\pm 1/2\rangle$. %, giving the matrix representation $\mathcal{M}_{x}=-i\tau_{z}\otimes\sigma_{x}$.
Using this representation, we find the eigenvalues $m_{z,1/2}=\pm i$ and $m_{z,3/2}=\pm i$, and the eigenvectors $\phi_{z,1/2}^{\pm i}=\left(\begin{array}{c c} \mathcal{Z}_{\pm i} & \vec{0} \end{array}\right)$ and $\phi_{z,3/2}^{\pm i}=\left(\begin{array}{c c} \vec{0} & \mathcal{Z}_{\mp i}\end{array}\right)$. Here, $\vec{\mathcal{Z}}_{i}=\left(1\quad 0\right)$, $\mathcal{Z}_{-i}=\left(0\quad 1\right)$ and $\vec{0}=\left(0\quad 0\right)$.
The Hamiltonian in the basis $\left\{\phi_{z,1/2}^{i}, \phi_{z,3/2}^{i}, \phi_{z,1/2}^{-i}, \phi_{z,3/2}^{-i}\right\}$ is a block diagonal matrix, 
\begin{equation}
\mathcal{H}_{Na_{3}Bi}=\left(\begin{array}{cc}
h_{z,i}(\vec{k}) & 0\\
0 & h_{z,-i}(\vec{k})
\end{array}\right), 
\label{Mz}
\end{equation}
where the blocks describing the mirror projected states reads \begin{equation}
h_{z,\pm i}(k_{x},k_{y})=\left(\begin{array}{cc}
-\varepsilon+\xi k^{2} & \ell_{\pm}(\vec{k})\\
\ell_{\mp}(\vec{k}) & \tilde{\varepsilon}-\tilde{\xi} k^{2}
\end{array}\right).
\end{equation}
Thus, a perturbation breaking the mirror symmetry introduces off-diagonal matrix elements in Eq.~\ref{Mz}. For instance, according to our \textit{ab initio} calculations, a perpendicular electric field can be effectively included as a Rashba-type term that couples the Hamiltonians $h_{z,i}$ and $h_{z,-i}$, i.e., $h_{R}=i\lambda_{R}k_{-}$ (See supplemental material).    

The mirror Chern number, $\mathcal{C}_{M}$, is defined in terms of the states at the mirror symmetry invariant $k$-points. For instance, in the 3D-TCI SnTe, $\mathcal{C}_{M}$ is computed using only the $k$-points at the intersection between the mirror plane and the 3D-BZ. Here, since the intersection between the BZ and the mirror plane is the BZ itself, all $k$-points are  mirror symmetry invariant. Thus, $\mathcal{C}_{M}$ is calculated through the Berry phase, $\Omega_{n}^{\pm i}(k_x,k_y)$, defined in the whole BZ~\cite{PhysRevLett.106.106802,Mokrousov2015}, i.e., $\mathcal{C}_{M}=(\mathcal{C}_{i}-\mathcal{C}_{-i})/2$, where $\mathcal{C}_{\pm i}$ Chern number for mirror eigenvalues $\pm i$,
\begin{equation}
\mathcal{C}_{\pm i}=\frac{1}{2\pi}\sum_{n<E_{f}} \int_{BZ} \Omega_{n}^{\pm i}(k_x,k_y) dxdy .
\end{equation}
Using the proposed model, we verified that $\mathcal{C}_{\pm i}= \mp 1$, indicating that the Na$_{3}$Bi monolayer is a TCI with $\mathcal{C}_{M}=-1$, as predicted in Ref.~\onlinecite{Na3Bi_2017}.

\textit{Prediction of DTC in HF-honeycomb materials:} Graphene-like materials functionalized with group-VII atoms are predicted to behave as QSHIs~\cite{2018arXiv180510950M}.
When a VII-atom is removed from the unit cell, one sublattice exhibits dangling bonds and hence, magnetic moments due to unpaired electrons possibly spontaneously evolve into a magnetic order, spoiling the QSH states and given rise to the quantum anomalous Hall effect~\cite{PhysRevLett.113.256401}. The electrons can be paired by substituting the non-passivated IV-atoms by atoms with an odd number of valence electrons. Thus, HF IV-V hexagonal-lattices materials, formed by two triangular sub-lattices, one consisting of an $\mbox{IV-VII}$ dimer and the other of atoms $\mbox{V}$ (See Fig \ref{f:Fig1}), have no magnetic moment. Therefore, this systems are predicted to be mechanically stable QSHIs~\cite{MeraAcosta2016}. 
%%%%%%%%%%%%%%%%%%%%%%%%%%5
Since the functionalization breaks the mirror symmetry $\mathcal{M}_{z}$, the symmetry operation satisfied by HF-compounds are only $\mathcal{R}_3$, $\mathcal{T}$, and $\mathcal{M}_{x}$, which is schematically represented in Fig. \ref{f:Fig1}b. 
In these systems, the TCI phase is protected by the reflection symmetry respect to the three planes that are perpendicular to the lattice and contain the lines connecting the nearest neighbors in the honeycomb lattice.

HF-honeycomb lattice materials can display an inverted band character dominated by Bi-atomic orbitals, like the Na$_{3}$Bi compound. Similarly, in the inverted order, the VBM has a $p$-orbital character, mainly dominated by $p_{z}$-Bi orbitals, whereas the CBm at $\Gamma$ mainly consists of $p_{x,y}$-Bi orbitals.
Hence, the Hamiltonian in the full SOC basis $\{|\textrm{Bi}_{1/2},\pm 1/2\rangle,|\textrm{Bi}_{3/2},\pm 1/2\rangle\}$ reads~\cite{MeraAcosta2016}
\begin{equation}
\mathcal{H}_{HF}=\left(\begin{array}{cccc}
-\varepsilon +\xi  k^{2}& i\alpha k_{-} & 0 & \gamma k_{-}\\
-i\alpha k_{+} & -\varepsilon +\xi k^{2}& \gamma k_{+} & 0\\
0 &  \gamma k_{-} &  \tilde{\varepsilon} -\tilde{\xi}  k^{2} & 0 \\
 \gamma k_{+} & 0 &0 & \tilde{\varepsilon} - \tilde{\xi} k^{2}
\end{array}\right),
\label{Halmiltonian}
\end{equation}
where $\alpha_{R}$ is the Rhasba parameter, and $ \gamma$ is the interaction term between the $J=1/2$ and $J=3/2$ states. 

%\begin{figure}
%  \centering
%  \includegraphics[width = 0.95 \linewidth]{Band}
%  \caption{(left) DFT calculation of the band structure with and without SOC for the compound PbBiI. \textit{Ab initio} calculations are performed within the density functional theory (DFT) framework as implemented in the SIESTA code\cite{soler2002siesta} with the on-site approximation for the SOC\cite{PhysRevB.89.155438,fernandez2006site}. The Perdew-Burke-Ernzenhof generalized gradient approximation\cite{PhysRevLett.77.3865} is used for the exchange-correlation functional. The SOC inverts the states with the total angular momentum $J=1/2$ and $J=3/2$, to which the $p_{z}$ (cian) and $p_{xy}$ (purple) orbitals from Bi mainly contribute, respectively.
%This band inversion takes place at the $\Gamma$ point. 
%The band gap is represented by gray areas. (right) Band structure calculated from Eq.~\ref{Halmiltonian} for $\vec{k}$-points in the mirror plane $M_{x}$ (blue plane). The bands are discriminated by the mirror symmetry operator eigenvalues $i$ and $-i$, which are represented in green and purple, respectively. The arrows at the energy cuts in the valence and conduction bands represent the in-plane spin texture.} 
 %\label{f:Fig2}
%\end{figure}

%Rashba spin-splitting leads to bands with opposite helical spin texture, 
%Spin-polarized Rashba semiconductor exhibit two parabolic bands with opposite helical spin texture,
The Hamiltonian describing QSHIs with inversion symmetry breaking (Eq.~\ref{Halmiltonian}) leads to parabolic bands with the same helical in-plane spin texture, forbidding the backscattering in bulk states~\cite{MeraAcosta2016}. This unconventional spin topology was also observed in the non-topological alloy Bi/Cu(111)~\cite{PhysRevB.79.245428}. %, different from the usual Rashba spin-splitting exhibiting opposite helical spin texture. 
We find that the scattering processes are not only limited by the spin texture. Specifically, since the commutation relation $[\mathcal{H}(k_{x}=0),\mathcal{M}_{x}]$, the wave-function can be indexed with the eigenvalues of the mirror operator $\mathcal{M}_{x}$, $m_{x,J}$, for $k_{x}=0$.
%This is a consequence of commutation relation $[\mathcal{H}(k_{x}=0),\mathcal{M}_{x}]$, which can be easily verified using the  matrix representation $\mathcal{M}_{x}=-i\tau_{z}\otimes\sigma_{x}$.,~i.e.,~the mirror operator transforms the orbitals as $\mathcal{M}_{x}|\textrm{Bi}_{1/2},\pm 1/2\rangle=-i|\textrm{Bi}_{1/2},\mp 1/2\rangle$ and $\mathcal{M}_{x}|\textrm{Bi}_{3/2},\pm 1/2\rangle=i|\textrm{Bi}_{3/2},\mp 1/2\rangle$. %, giving the matrix representation $\mathcal{M}_{x}=-i\tau_{z}\otimes\sigma_{x}$.

\noindent In the basis $\left\{\phi_{x,1/2}^{i}, \phi_{x,3/2}^{i}, \phi_{x,1/2}^{-i}, \phi_{x,3/2}^{-i}\right\}$, the Hamiltonian is a block diagonal matrix, 
\begin{equation}
\left.\tilde{\mathcal{H}}(\vec{k})\right|_{k_{x}=0}=\mathbbm{1}\otimes h_{0}+\tau_{z}\otimes(h-h_{R}),
%\left(\begin{array}{cc}
%h(k_{y}) & 0\\
%0 & h(-k_{y})
%\end{array}\right), 
\label{unitary}
\end{equation}
where $h=\sigma_{y}\gamma k$ and $h_{R}=(\mathbbm{1}+\sigma_{z})\alpha k/2$, and
\begin{equation}
h_{0}(k)=\left(\begin{array}{cc}
-\varepsilon+\xi k^{2} & 0\\
0 & \tilde{\varepsilon}-\tilde{\xi} k^{2}
\end{array}\right).
\end{equation}
The blocks of the Hamiltonian related to the eigenvalues of the mirror symmetry operator $\pm i$ are written as $h_{\pm i}=h_{0}\pm (h-h_{R})$, respectively. The Hamiltonian $h_{\pm i}$ leads to the band structure discriminating mirror eigenvalues.
Since the intersection between the mirror plane and the BZ is a line (See Fig.~\ref{f:Fig1}), the mirror Chern number is defined in terms of the one-dimensional $k$-points at $k_{x}=0$. Specifically, $\mathcal{C}_{\pm i}$ are calculated via the winding numbers~\cite{PhysRevB.89.125425}, which is essentially the Zak phase for the one-dimensional effective Hamiltonian $h_{\pm i}$~\cite{Winding_numbers1D,Zak,Exp_Zak}.
Remarkably, in the Hf-honeycomb material, the edge states are protected by the mirror plane perpendicular to the structure ($\mathcal{M}_{x}$), which is different from the conventional TCIs. This offers the possibility of breaking both TR and mirror symmetries with a perpendicular magnetic field (parallel to the mirror plane $\mathcal{M}_{x}$) and also allows to manipulate the edge states with in-plane magnetic fields (See supplemental material).
In the general case, the magnetic field $\mathcal{H}_{B_{\parallel\mathcal{M}_{x}}}=\mathbbm{1}\otimes(\sigma_{y}B_{y}+\sigma_{z}B_{z})$, gives rise to a coupling term between $h_{i}$ and $h_{-i}$, inducing a spin-polarization in the direction in which is applied, even for $k$-point at the mirror plane.

\textit{Electric field effect in 2D-materials with DTC:}
\begin{figure}
  \centering
  \includegraphics[width = 0.95 \linewidth]{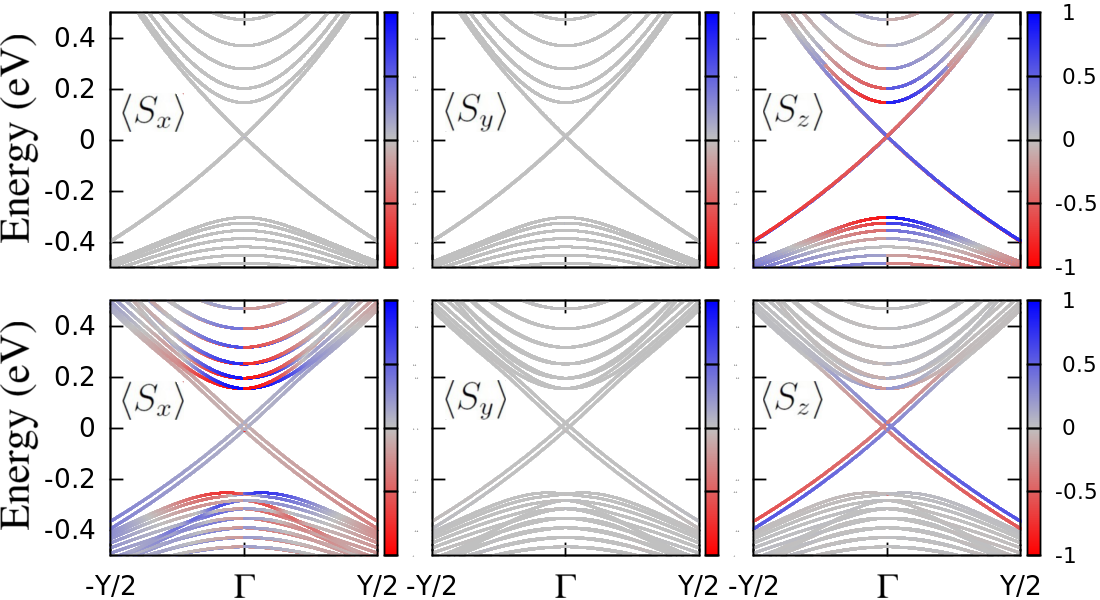}
  \caption{Expected values of the spin operator for the band structure of a Na$_3$Bi nanoribbon 37.1~nm width armchair terminated for (top) $E_{0}=0\mbox{ V/\AA}$ and (bottom) $E_{0}=0.1\mbox{ V/\AA}$. Here, $E_{0}$ is an external electric field breaking the mirror symmetry $\mathcal{M}_{z}$. The color code stands for the spin orientation.}
 \label{f:Fig5}
\end{figure}
Metallic edge states, the most interesting feature in both TCIs and QSHIs, are computed by considering open boundary conditions in the previously discussed tight-binding models.
First, we confirm the presence of edge states preserving the TR symmetry, i.e., anti-propagating spin current in each edge. Before including the electric field effects, the spin is forced to be perpendicular to the mirror plane to preserve this symmetry, i.e., oriented along the $z$-axis ($x$-axis) in Na$_{3}$Bi (HF-hexagonal) compounds, as shown in Fig.~\ref{f:Fig5} and~\ref{f:Fig4}.  
%Different from the out-plane spin polarized edge states in the known QSHIs, e.g., Na$_{3}Bi$, 
This suggests that external perturbations breaking the mirror symmetry could induce spin-polarization components parallel to the mirror plane, as we show below by considering an external electric field.

In the Na$_{3}$Bi compound, a perpendicular electric field breaks the symmetry $M_{z}$, as previously discussed. This field induces a helical spin-texture in the bulk band structure, leading to in-plane spin-polarization components in the edge states, i.e., parallel to $M_{z}$, as represented in Fig.~\ref{f:Fig5}. The edge states are still protected by the TR-symmetry and the electric field provides a mechanism to change the spin-polarization orientation.

An external electric field breaking the mirror symmetry $\mathcal{M}_{x}$ can be introduced by modifying the on-site term,~i.e.,~$\tilde{\varepsilon}_{n}(\vec{k})=\varepsilon(\vec{k})+ naeE_{0}/N$, where $n$ is the index of the Bi atoms in a nanoribbon formed by $N$ unit cells along the $x$-axis, $a$ is the lattice constant, and $e$ is the electron charge. We find that this field induces a non-zero out-plane spin-polarization, as represented in Fig.~\ref{f:Fig4}. 
%Although the Dirac cones are splitting to each other, the TR-symmetry is not broken. 

\begin{figure}
  \centering
  \includegraphics[width = 0.95 \linewidth]{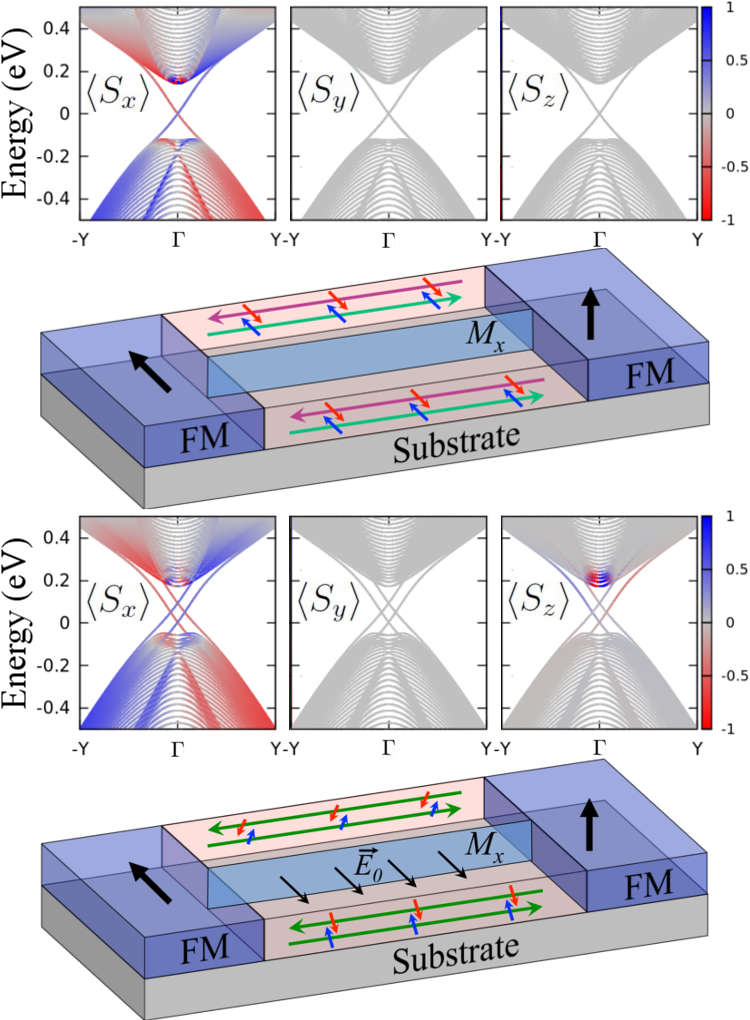}
  \caption{Expected values of the spin operator for the band structure of a nanoribbon 33.6~nm width armchair terminated for (top) $E_{0}=0\mbox{ V/\AA}$ and (bottom) $E_{0}=2.67\cdot 10^{-2}\mbox{ V/\AA}$. Here, $E_{0}$ is an external electric field breaking the mirror symmetry $\mathcal{M}_{x}$ (blue plane). The color code stands for the spin orientation. The spin texture is also schematically represented in a transistor model: a nanoribbon connected to two ferromagnetic contacts (FM). This structure is deposited in a substrate (gray area). States with eigenvalue symmetry operator $\pm i$, which are only defined if the electric field is zero, are represented by purple and green, respectively.}
 \label{f:Fig4}
\end{figure}
The effect of an electric field breaking the mirror symmetry can be understood using the phenomenological model describing the Rashba effect in inversion symmetry breaking systems: 
the electric field $\vec{E}$ induces a momentum-dependent Zeeman energy $\vec{\phi}_{eff}=\mu_{B} \vec{\mathcal{B}}_{eff}\cdot\sigma$ with $\vec{\mathcal{B}}_{eff}=\vec{E}\times\hbar\vec{k}/m_{e}c^{2}$~\cite{Rashba1984,Manchon2015}. 
This odd-in-$k$ effective field, i.e.,~$\vec{\mathcal{B}}_{eff}(k)=-\vec{\mathcal{B}}_{eff}(-k)$, preserves the Tr-symmetry and only appears when the mirror symmetry is broken, in the same way that the Rashba effect depends on the inversion symmetry breaking~\cite{Manchon2015,Vajna2012}.
%The odd dependence on $k$ imposes that the total spin-polarization component vanishes, preserving the TR-symmetry, which is consistent with the spin-polarization observed in Fig.~\ref{f:Fig5} and~\ref{f:Fig4}. 
Analogous to the generic helical edge states~\cite{PhysRevB.93.205431,1367-2630-12-6-065012}, this field changes the direction of the spin polarization, as observed in Fig.~\ref{f:Fig5} and~\ref{f:Fig4}.
%This effect has been widely studied in spin-polarized Rashba semiconductors~\cite{SpinCurrent2012}. 
Specifically, in arm-chair nano-ribbons, the momentum carried by the electrons at the edge is oriented along the $y$-axis. In HF-honeycomb (Na$_{3}$Bi) compounds, the external electric field is transverse (perpendicular) to the nanoribbon $\vec{E}=E_{0}\hat{x}$ ($\vec{E}=E_{0}\hat{z}$), resulting in an effective field along the $z$-axis ($x$-axis), $\mathcal{B}_{eff}=(\mu_{B}/m_{e}c^{2})E_{0}k_{y}$.

The mirror symmetry protection in HF-hexagonal materials implies that states spatially localized in different edges have the same momentum and spin direction. 
Thus, the spin-flip is required in scattering processes involving these states. 
This spin texture has also been observed in curve QSHIs~\cite{ncomms15850}.
In zigzag nanoribbons, the mirror symmetry is intrinsically broken and hence, the spin is not only oriented along the $x$-axis~\cite{MeraAcosta2016}. %, but also in the $z$-axis. 
Although giant Rashba-splitting coexists with surface states in
the 3D TIs due to the band bending and structural inversion asymmetry~\cite{ncomms1131,PhysRevLett.107.096802,PhysRevB.88.081103}, the consequences of the Rashba effect and transverse electric fields in surface states have not been widely explored~\cite{APL3664776}. Based on the tight-binding model proposed in Ref.~\onlinecite{PhysRevB.98.035106}, we verify that spin-polarization can also be controlled in 3D TIs with DTC (See supplemental material).
The spin-polarization control proposed in this letter is different from the dynamic spin-orbit torque in anti- and ferromagnetic 2D and 3D TIs~\cite{nmat2613,PhysRevB.95.035422,PhysRevLett.114.257202,NatNanoJungwirth2014,PhysRevB.91.134402}, since the effective field $\vec{\mathcal{B}}_{eff}$ appears for the equilibrium carrier spin density  configuration and does not require magnetic order.

The change of spin orientation suggests that a simple transistor model can be constructed. 
If the armchair nano-ribbon of HF-honeycomb materials (Na$_{3}$Bi) is connected to ferromagnetic electrodes, the electrons in the edge states are not detected by a drain whose magnetic moment is perpendicular to the $x$-axis ($z$-axis), as schematically represented in Fig.~\ref{f:Fig4}. This corresponds to the Off of the transistor. If an electric field perpendicular to the mirror plane is turned on, the electrons in the edge states have a non-zero probability of being detected by the electrode (See Fig.~\ref{f:Fig4}). Different from the transistors based on TCIs, in DTC insulators, the switch is not defined by the band-gap opening, but the spin-polarization direction. 

%Since the PbBi alloy can be experimentally realized keeping the rhombohedral structure~\cite{Gokcen1992,Huang1983}, this structure can be in turns the substrate
%the Bi-Pb alloy can be realized experimentally maintaining the $R\bar{3}m$ space group \cite{Gokcen1992,Huang1983}. The Pb-Bi rhombohedral alloy along the [111] direction can be considered as a stack of PbBi honeycomb lattices that are weakly bonded (mainly ruled by Van der Waals type interaction) to each other, similarly to the bismuth bilayers\cite{Drozdov2014}. The dangling bonds that appear at the Pb-rich PbBi surface can be eliminated by bonding to iodine atoms and hence, the proposed spin texture could be observed in the PbBiI system via STM experiments analogously to the observation of Bi-bilayers' edge states\cite{Drozdov2014}.

Summarizing, we proposed a transistor model using the spin-direction as a switch. We showed that in systems simultaneously protected by the TR and mirror symmetries, the spin-polarization is always perpendicular to the mirror plane protecting the metallic edge states. We demonstrated that an external electric field breaking the mirror symmetry induces a spin-polarization parallel to the mirror plane. 
Since the electric field does not break the TR-symmetry, the metallic edge states are still protected by this symmetry, which allows the control of the spin-polarization preserving the Dirac cone formed by these states.
We constructed tight-binding models for Na$_{3}$Bi and HF-honeycomb materials. Using this model we computed the mirror Chern number, predicting the DTC in HF-honeycomb materials and verifying this behavior in the Na$_{3}$Bi compound, which are protected by the mirror plane $\mathcal{M}_{x}$ and $\mathcal{M}_{z}$, respectively. In order to illustrate that the proposed spin-polarization control does not depend on the mirror plane orientation, we study this effect in both materials, the only two-dimensional systems predicted to exhibit a DTC. In light of recent advances in the application of electric fields in the 2D Na$_3$Bi~\cite{NatureCollins2018}, the practical realization of the proposed device is feasible. 

%Since they are protected by different mirror planes, the proposed spin-polarization control 
%We also showed that the spin-texture of the egde states is alway perpendicular to the mirror plane protecting them and hence, the spin-polarization orientation change by breaking this symmetry. 

%An in-plane magnetic fields $B=B_{x}\hat{x}$, breaking the TR-symmetry and preserving the mirror symmetry $M_{x}$, leads to a displacement of the Dirac point formed by the edge states in the reciprocal space, also confirming the topological mirror symmetry protection. Additionally, band-gap sizes engineering can be performed by breaking both the TR and mirror symmetries by applying an external magnetic field parallel to the mirror plane. 

\begin{acknowledgements}
This work was supported by the Sao Paulo research foundation (grants 2014/12357-3 and 17/02317-2). We would like to thank
Dr. Gerson J. Ferreira and Dr. Marcio Costa for the discussions.
\end{acknowledgements} 

%\begin{bibliography}
\bibliography{Ref}
%\bibliographystyle{phaip}
%\end{bibliography}

\end{document}